\documentclass[conference]{IEEEtran}
\IEEEoverridecommandlockouts
\usepackage{cite}
\usepackage{amsmath,amssymb,amsfonts}
\usepackage{algorithmic}
\usepackage{graphicx}
\usepackage{textcomp}
\usepackage{xcolor}
\usepackage{url}
\usepackage[hidelinks]{hyperref}
\usepackage{graphicx,alltt,xcolor,float}
\usepackage{listings}
\usepackage{comment}
\def\BibTeX{{\rm B\kern-.05em{\sc i\kern-.025em b}\kern-.08em
	T\kern-.1667em\lower.7ex\hbox{E}\kern-.125emX}}
\begin{document}

\title{VoIP Can Still Be Exploited --- Badly}

\author{
	\IEEEauthorblockN{Pietro Biondi, Stefano Bognanni and Giampaolo Bella}
	\IEEEauthorblockA{\textit{Dipartimento di Matematica e Informatica}\\
		\textit{Universit\`a di Catania}\\
		Catania, Italy\\
		pietro.biondi@phd.unict.it, stefano.bognanni97@gmail.com, giamp@dmi.unict.it}
}

\maketitle

\begin{abstract}
	VoIP phones are early representatives as well as present enhancers of the IoT. This paper observes that they are still widely used in a traditional, unsecured configuration and demonstrates the Phonejack family of attacks: Phonejack 1 conjectures the exploitation of phone vulnerabilities; Phonejack 2 demonstrates how to mount a denial-of-service attack on a network of phones; Phonejack 3 sniffs calls. It is reassuring, however, that inexpensive devices such as a Raspberry Pi can be configured and programmed as effective countermeasures, thus supporting the approach of integrating both technologies. We demonstrate both attacks and defence measures in a video clip \cite{videophonejack}. The concluding evaluations argue that trusting the underlying network security measures may turn out overly optimistic; moreover, VoIP phones really ought to be protected as laptops routinely are today.
	
\end{abstract}
\section{Introduction}\label{sec:intro}

The IoT loosely refers to any interconnected devices used to monitor and control information in order to deliver services to remote users. Examples of such devices are countless, including for example smart plugs and lights, coffee makers and heating systems. The list is growing steadily, and we feel that the IoT era is only just dawning \cite{introIoT}. 

Wherever useful services are available, malicious attacks arouse.
One of our favourite examples is the Samsung refrigerator leveraged to violate a set of Gmail credentials \cite{smartfridge}. Another example we contributed is about printers, whose unprotected 9100 ports allow an attacker to mount a paper DoS as well as to eavesdrop the contents of the printouts \cite{overtrustprinter}. 

VoIP devices and related protocols, such as Session Initiation Protocol (SIP) \cite{SIP} and Real-time Transport Protocol (RTP) \cite{RTP}, revolutionised traditional voice calling technology. They brought up at the level of computer networks a service that traditionally run at another, separate level. Resting on a long-prototyped technology, the popularity of VoIP dates back to approximately the mid 1990s \cite{historyvoip}, hence VoIP devices can be considered among the earliest IoT members. Moreover, their integration with the current IoT is gaining significant momentum as we write, also due to the quest for controlling devices and services remotely by voice \cite{azienda1,azienda2,azienda3}.

The weaknesses of VoIP in front of malicious attackers are known at least since the ``Information Security Reading Room'' of the SANS Institute published an eminent report in 2002 \cite{sans}. The report provided a proof-of-concept of how VoIP calls could be overheard by using \textit{commercial tools}.  Our research aims at verifying whether and how that work has been universally received today --- after nearly two decades --- namely whether VoIP in use has been hardened at all. Our methodology is empirical and leverages \textit{freeware} to conduct a Vulnerability Assessment and Penetration Testing session on the VoIP devices currently in use in our Department. The outcome is that those devices are variously exploitable from inside the Departmental network, although it is clear that the network has protection measures from the outside. We are aware that the very same devices are adopted in a number of other Institutions under similar configurations, so the same outcome could be expected elsewhere too.

VoIP hardening measures exist today, such as TLS-based solutions SIPS \cite{SIPS} and SRTP \cite{SRTP}. However, our findings demonstrate that \textit{there are devices still in use at present} that are as weak as two decades ago. This may be interpreted as yet another paradox of security economics \cite{ross}, but particularly surprises us for at least two reasons.
One is that the use of VoIP technology is widespread and, as noted above, additionally empowered lately. 
The other one is that protection measures at the network level have known limitations because abuse at node level is still possible. Broader evaluations will come at the end (\S\ref{sec:concl}).

\subsection{Contributions}\label{sec:contrib}
This paper explores whether and how VoIP devices can be exploited today using freeware, namely non-commercial tools that teenagers may try out since school.
We address this question in a specific though common scenario: phones do not run TLS-based solutions; an insider attacker connects her attacking laptop to an Ethernet cable unplugged from a VoIP phone. The findings are that, based upon simple tools such as nmap, Ettercap, Wireshark and Python programming, the attacker can seriously compromise the VoIP service.

More precisely, we define a family of three attacks and, following the same style used against printers before \cite{overtrustprinter}, we term it the \textit{Phonejack family of attacks against VoIP}:
\begin{itemize}
	\item \textbf{Phonejack 1 attack: Zombies for DDoS.} Every model of a VoIP device may suffer zero-day or documented vulnerabilities. These could be exploited to carry out large-scale DDoS attacks against specific Internet targets. We conjecture but do not attempt such attack (\S\ref{sec:zombie}).
	\item \textbf{Phonejack 2 attack: phone DoS.} By continuously sending packets to a target VoIP device, this can be exploited to ring indefinitely and crash eventually. We leverage Python multi-thread programming to overwhelm a test network of four VoIP devices and publish a video clip to demonstrate the audio experience of the attack (\S\ref{sec:voipdos}). To the best of our knowledge, Phonejack 2 is entirely innovative.
	\item \textbf{Phonejack 3 attack: audio call eavesdropping.} Because packets are sent in the clear, we eavesdrop successfully and dump them to an audio file (\S\ref{sec:privacyattack}). This attack is based on the mentioned 2002 SANS observations \cite{sans} but it is intriguing that it solely relies on freeware.
\end{itemize}

We also prototype effective countermeasures, at least against Phonejack 2 and 3. Taking advantage of inexpensive Raspberry Pi devices, every VoIP phone can be shielded to only accept each distinct traffic packet once, hence countering Phonejack 2. Raspberry Pis can also implement encryption so that packets are only transferred enciphered over the network hence cannot be understood by a man in the middle (MITM). Each Pi then relays cleartext packets to its phone. While this prototype somewhat optimistically trusts the network between a Pi and its phone, it demonstrates once more how a security measure could be incorporated and deployed in a VoIP phone.

Our experiments were conducted over an air gapped testbed detailed below. 
Taking inspiration from them, we took the mere information gathering step at institutional level, finding out that the phone network and the computer network were not adequately separated. We reported this to our IT team, and corrective measures were implemented immediately.

\subsection{Testbed}
Our test-bed is an air-gapped network featuring an Asterisk server version 16, an attacking laptop running the offensive freeware, then one VoIP phone model Cisco SPA 921 and three phones model Cisco SPA 922. All are connected through a Netgear gs105se switch. Alternatively, because Cisco SPA 922 features two Ethernet sockets, one for incoming and one for outgoing traffic, the attacking laptop could be connected to one such phone. 

The additional devices used to demonstrate the use of encryption to protect the calls were a Raspberry Pi 3 b+ e and a Pi 4 b. These were connected through a different network setup: each Pi to a phone via Ethernet, then the Pis to each other via their Wi-Fi interface through a SSID exposed by a Vodafone Station Revolution.

\subsection{Paper Structure}
This manuscript continues describing our Phonejack family of attacks (\S\ref{sec:zombie},\S\ref{sec:voipdos},\S\ref{sec:privacyattack}) and some possible countermeasures  (\S\ref{sec:countermeasures}). It outlines some related work (\S\ref{sec:related}) and concludes with some broader evaluations of the findings (\S\ref{sec:concl}). The basics of the underlying VoIP protocols SIP and RTP are deferred to Appendix due to space constraints.

\section{Phonejack 1 attack: Zombies for DDoS} \label{sec:zombie}
Denial of Service (DoS) is one of the most dangerous attacks. A simple example would be a large infrastructure that no longer responds to millions of users  for hours and hours. The distributed version of denial of service (DDoS) floods the victim with traffic generated from different sources. To perform a DDoS, an attacker builds a botnet, a network of infected machines called zombies. The prolonged duration of a DDoS can cause may have various logistic and monetary consequences, such as loss of customer trust towards the service and monetary damage to the business up to \texteuro 35.000 per hour over an average duration of 15 hours\cite{DDoS1}. In consequence, when new vulnerabilities are discovered on remotely controlled IoT devices, there is also a danger of exploiting these devices as zombies. So, even if the computational power of the single device is not significantly high, the huge number of available devices form together a valuable computing asset. For example, a botnet of 1.5 million cameras was built in 2016 and generated 660 Gbps of network traffic \cite{camerasvice}.

Therefore, we seek out to assess what vulnerabilities are known of VoIP systems. Querying the Common Vulnerabilities and Exposures (CVE) database of the MITRE \cite{homemitre} with keyword ``voip'' returns 107 CVE entries, some of which can be practically exploited on devices \cite{cvevoip}. The query can then be specified over the devices in our setup (\S\ref{sec:contrib}) yielding two CVEs, namely CVE-2014-3312, a Remote Command Execution (RCE) \cite{CVE-2014-3312}, and CVE-2014-3313, a Cross-Site Scripting (XSS) \cite{CVE-2014-3313}.

The impact of exploiting such vulnerabilities could be evaluated by considering the DDoS implications mentioned above. Querying Shodan \cite{shodan} may, in turn, help us assign a likelihood to such vulnerabilities, by informing us of how common the affected devices are. A ``cisco spa'' query returns 455 entries, which is a rather low outcome. A possible explanation is that VoIP devices are not left publicly visible, which is a commendable protection measure. However, a more general query, say ``asterisk'', returns 59.341 results, and would give a motivated attacker thousands of potential targets worth of further vulnerability assessment to seek VoIP exploitation. Of course, 2014 vulnerabilities have arguably been fixed ever since, but then we question whether updating phones falls into widespread security maintenance routine. We refrain from actively engaging into exploiting such vulnerabilities because this lies outside our research aims.

Finally, it must be recalled here that VoIP phones may also suffer undocumented, zero-day attacks.

\section{Phonejack 2 attack: phone DoS} \label{sec:voipdos}
While the exploitation of VoIP phones in a botnet might be considered somewhat ``traditional'', we also wonder to what extent phones themselves can become victims of a DoS activities. To assess such a vulnerability, we explore how to bombard a phone with tailored SIP packets, and observe that this can be successful. 

As preliminary operations, we configure the Asterisk server and the four Cisco VoIP phones so that they authenticate to the server. A phone ID, phone number, password and gateway must be manually entered on each phone. We assume that the phone number is public. An attacker may build a function to scan the local network and obtain the IPs and MAC addresses of the connected devices. Table \ref{tab:netscanscript} shows a Python implementation of such a function. The \texttt{network} parameter represents the network to be scanned (e.g. 192.168.1.0/24).
\begin{table}[h]
\caption{Network scanning in Python}\label{tab:netscanscript}
\begin{lstlisting}
def scanNetwork(network):
 hosts = []
 nm = nmap.PortScanner()
 out = nm.scan(hosts=network, arguments='-sP')
 for k, v in out['scan'].iteritems():
  if str(v['status']['state']) == 'up':
   hosts.append([str(v['addresses']['ipv4']),
   str(v['addresses']['mac'])])
return hosts 
\end{lstlisting}
\end{table}
We now make a call between two phones and record the network traffic through Wireshark \cite{wireshark}, as if run by an attacker. We apply a filter to extract the SIP packet that causes a ring, and save this packet in a file called \texttt{sipInvite.pcap}. This file contains information such as number and IP address of the recipient phone. We note that a phone does not check the SIP timestamp of a received packet but only that the recipient phone number of the packer corresponds to itself. Thus, a receiving phone only checks whether a packet is intended for itself. We also observe that flooding a phone with requests causes it to ring continuously, then crash and reboot. These observations guide our attack. We write a function (again in Python) to carve a SIP packet as we want. Our \texttt{flood\_DoS} function in Table \ref{tab:floodDoS} takes an id, an IP address and a MAC address and calls the \texttt{tcprewrite} and \texttt{tcpreplay} commands. More specifically, \texttt{tcprewrite} takes the \texttt{sipInvite.pcap} file as input and modifies the fields containing the IP and MAC address of the packet. Finally, \texttt{tcprewrite} saves the new forged phone-ringing packet in a file called \texttt{newSipInvite.pcap}. After that, the \texttt{tcpreplay} command takes the \texttt{newSipInvite.pcap} file, sends it in loop to the phone and achieves the expected outcome: the phone rings for a few seconds, then crashes and reboots.
\begin{table}[h]
\caption{Phonejack 2 attack in Python}\label{tab:floodDoS}
\begin{lstlisting}
def flood_DoS(id, IP, MAC):
 subprocess.call([`tcprewrite',
 `--dstipmap=192.168.1.18:'+IP,
 `--enet-dmac='+MAC,`--dlt=enet',`--fixcsum',
 `--infile=sipInvite.pcap',
 `--outfile=newSipInvite'+id+`.pcap'])

 subprocess.Popen([`tcpreplay', `--intf1=eth0',
 `--loop=5',`newSipInvite'+id+`.pcap']
return
\end{lstlisting}
\end{table}

We then to carry out this attack in parallel on all our four devices. Table \ref{tab:mainpy} shows a Python script that uses a thread for each phone. More precisely, the script scans the network using the \texttt{scanNetwork} function, builds a thread, gives it a job by means of the \texttt{flood\_DoS} function and starts it.
\begin{table}[h]
\caption{Parallelising the Phonejack 2 attack in Python}\label{tab:mainpy}
\begin{lstlisting}
if __name__ == ``__main__":
 hosts = scanNetwork(sys.argv[1])
 jobs = []
 for i in range(0, len(hosts)):
  IP=hosts[i][0]
  MAC=hosts[i][1]
  thr = threading.Thread(target=flood_DoS(i,IP,MAC))
  jobs.append(thr)
 for j in jobs:
  j.start()
 for j in jobs:
  j.join()
\end{lstlisting}
\end{table}

\begin{figure}[h]
	\begin{center}
		\includegraphics[scale=0.15]{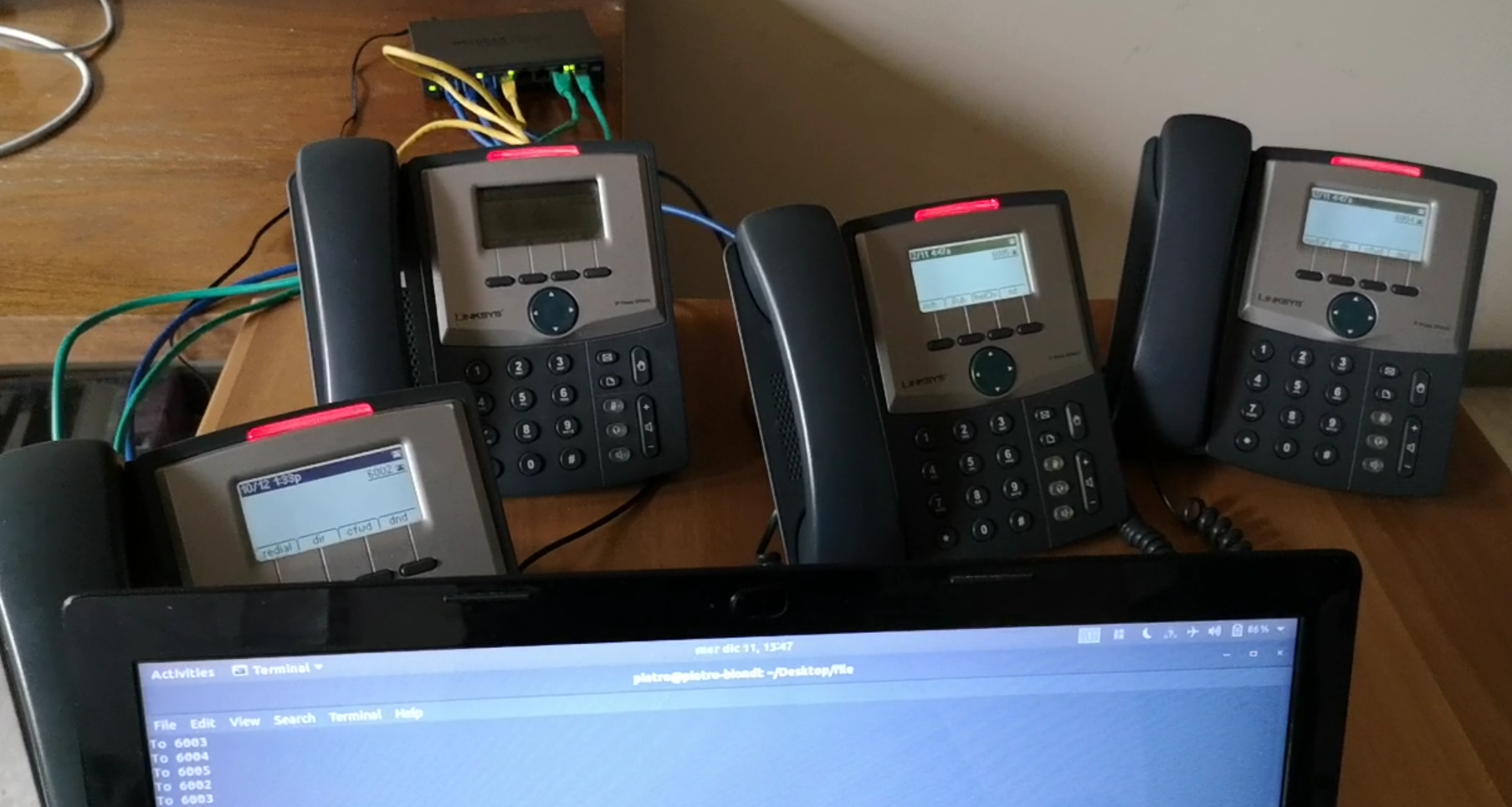}
	\end{center}
	\caption{Consequences of Phonejack 2}\label{fig:telphonejack2}
\end{figure}

Figure \ref{fig:telphonejack2} shows a laptop executing the Phonejack 2 attack against our four VoIP phones. Remarkably, all phones are ringing, as demonstrated by the red light on each of them. To better explain our results, we built a video clip \cite{videophonejack}. 

It can be imagined that mounting this attack on a departmental or institutional scale would have dramatic consequences. Not only would the calling capability dwarfed and ultimately zeroed, but the work environment would realistically become unbearable. We have not, however, scaled up our experiments.

\section{Phonejack 3 attack: audio call eavesdropping} \label{sec:privacyattack}

Let us assume that Alice and Bob want to get in touch and that the attacker Eve is present in the same network. Since calls are made in the clear, Even could attempt sniffing a call between Alice and Bob, clearly infringing their privacy. 

\begin{figure}[h]
	\begin{center}
		\includegraphics[scale=0.27]{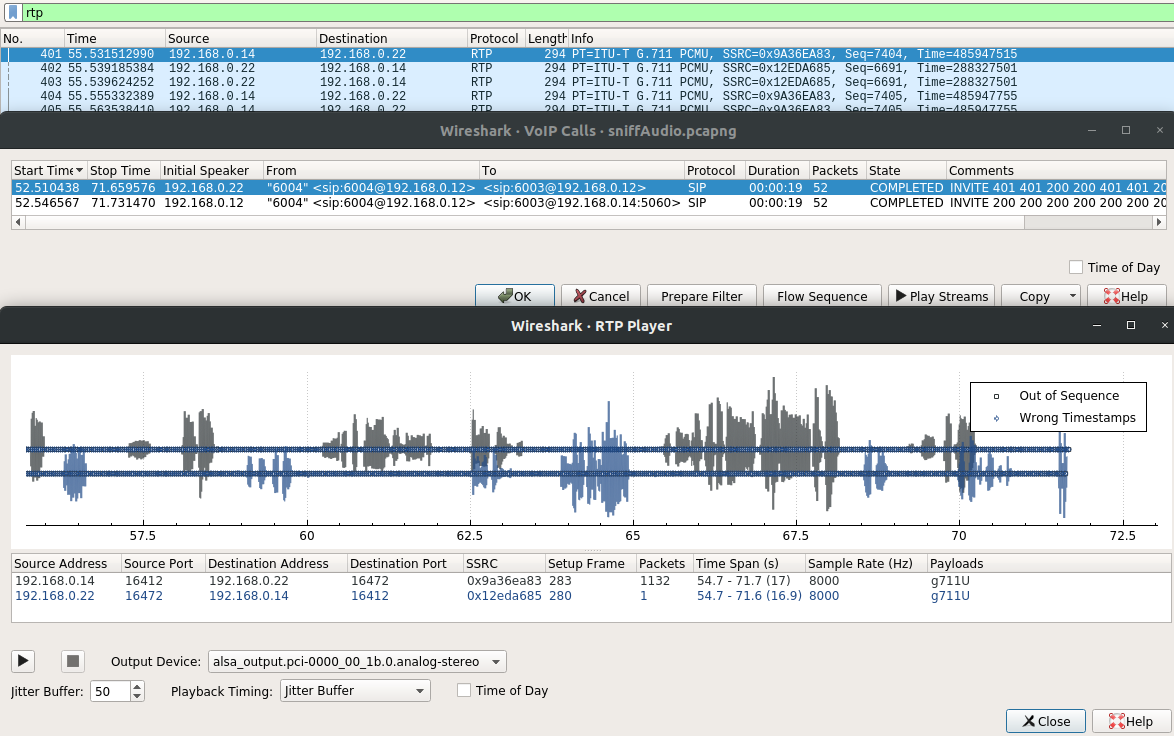}
	\end{center}
	\caption{Consequences of Phonejack 3}\label{fig:phonejack3}
\end{figure}

This conjecture can be demonstrated by taking the following steps. First, use Ettercap \cite{ettercap} to perform a Man in The Middle attack. Then, use a feature of Wireshark to listen to the audio flow of communication between two devices. Figure \ref{fig:phonejack3} shows the RTP traffic in the clear, as sniffed through Wireshark and played. To do this, we select an RTP packet, use the \texttt{Telephony} option and select the \texttt{VoIP Calls} feature. After that, we select one of the two streams and press \texttt{Play Stream}. Moreover, at the end of the call, we can export the the audio track of the call as shown in our video clip \cite{videophonejack}.

Clearly, this attack could be leveraged to exfiltrate data also at industrial espionage level. As noted above, it is not conceptually innovative but it is remarkable that we succeeded in carrying it out by using only freeware.

\section{Countermeasures}\label{sec:countermeasures}
We also design and develop countermeasures for the Phonejack 2 and Phonejack 3 attacks. This means that the countermeasures should thwart the malicious crashing of the phones as well as call sniffing.

\begin{figure}[h]
	\begin{center}
		\includegraphics[scale=0.3]{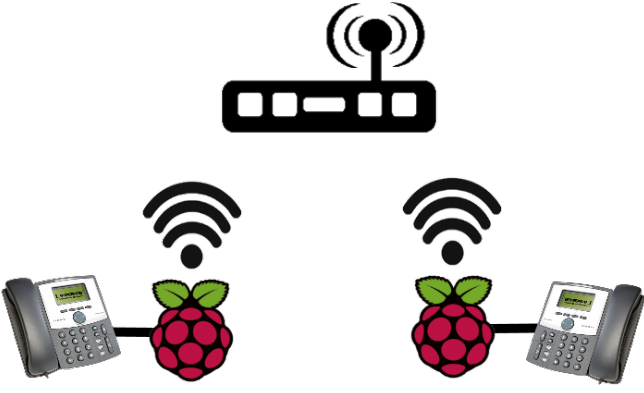}
	\end{center}
	\caption{An inexpensive network upgrade in support of our countermeasures}\label{fig:netArch}
\end{figure}

We opt for adopting an inexpensive device that could be easily programmed to prototype our solutions, and decide to leverage a recent model of Raspberry Pi. Two VoIP phones can be connected trough two Pis and a Wi-Fi bridge \cite{piwifibridge} as shown in Figure \ref{fig:netArch}. Precisely, each phone is connected to a Raspberry Pi through a wired Ethernet connection. Each Raspberry Pi will (have to) communicate via Wi-Fi with the other Raspberry Pi (because a Pi only has one Ethernet card).
For simplicity, we set static IP addresses by means of \textit{dhcpcd} and \textit{dnsmasq} tools. We initially connect the Raspberry Pi to the Wi-Fi router, then we modify the \textit{``/etc/dhcpcd.conf"} file by setting the interface, static IP address and the subnet. Then, we modify the \textit{``/etc/dnsmasq.conf"} file to tell dnsmasq how it should handle traffic. After that, we activate the forwarding mode of the network card and configure Iptables as shown in Table \ref{tab:iptablesrulenet} effectively bridge Ethernet and Wi-Fi. Finally, we update the routing tables of each Pi. 

\begin{table}[h]
\caption{Bridging Ethernet and Wi-Fi through Iptables}\label{tab:iptablesrulenet}
\begin{lstlisting}
iptables -t nat -A POSTROUTING -o wlan0
-j MASQUERADE

iptables -A FORWARD -i wlan0 -o eth0 -m state
--state RELATED,ESTABLISHED -j ACCEPT

iptables -A FORWARD -i eth0 -o wlan0 -j ACCEPT 
\end{lstlisting}
\end{table}

Having upgraded the network, we can turn out attention to the actual attack countermeasures. Since we need to modify the VoIP flow, we use Iptables to redirect traffic and build 3 queues with which the three scripts will be associated (Table \ref{tab:queues}). Specifically, queue 1 will be assigned with a dedicated anti-DoS script, then queues 2 and 3 respectively with scripts to encrypt and decrypt audio traffic.

\begin{table}[h]
\caption{Enqueuing SIP and RTP traffic through Iptables}\label{tab:queues}
\begin{lstlisting}
1) iptables -A FORWARD -p UDP -d PhoneAddress
--dport 5060 -j NFQUEUE --queue-num 1

2)iptables -A FORWARD -p UDP -s IPPhoneAddress
--sport rangeRTPport -j NFQUEUE --queue-num 2  

3)iptables -A FORWARD -p UDP -d PhoneAddress
--dport rangeRTPport -j NFQUEUE --queue-num 3   

\end{lstlisting}
\end{table}

\subsection{Countering Phonejack 2}
Table \ref{tab:scriptantidos} shows our script to counter Phonejack 2. It analyses each packet received via the \texttt{get\_payload} function. It checks the file called \texttt{blacklist.txt}. If the analysed packet has been previously received, then it is discarded, otherwise it is accepted and marked as received in the file. The penultimate statement builds a queue via the \texttt{NetfilterQueue} library, while the last instruction connects the ID and the anti-DoS script to queue 1.
\begin{table}[h]
\caption{A Phonejack 2 countermeasure in Python}\label{tab:scriptantidos}
\begin{lstlisting}
def antiDos(packet):
 pkt = IP(packet.get_payload())
 Flag=0
 with open('blacklist.txt') as f:
  if str(packet.get_payload()) in f.read():
   Flag=1
   if Flag == 1:
   packet.drop()
  else:
  packet.accept()
  f= open("blacklist.txt","a+")
  f.write(str(packet.get_payload()))
  f.close()
  Flag=0
nfqueue = NetfilterQueue()
nfqueue.bind(1, antiDos)
\end{lstlisting}
\end{table}

At this point, as shown in our video clip \cite{videophonejack}, each Raspberry Pi acts as a shield for a phone by filtering old packets from new ones while preserving voice communication.

\subsection{Countering Phonejack 3}
In this section we implement our solution to encrypt and decrypt the audio stream without any significant overhead or additional latency to the call. Table \ref{tab:scriptenc} shows the encryption script. It runs a few preliminary cryptographic operations and then executes the encryption function on the packet payload. Subsequently, it sends the encrypted packet via a socket and dequeues the packet. As with the anti-DoS script, also this script also be associated with a queue, in this case with the queue 2.
\begin{table}[h]
\caption{A Phonejack 3 countermeasure in Python}\label{tab:scriptenc}
\begin{lstlisting}
def encrypt(packet):
 cipher_suite = Fernet(key)
 enc_vc=cipher_suite.encrypt(packet.get_payload())
 pkt = IP(packet.get_payload())
 MESSAGE = enc_vc
 sk = socket.socket(socket.AF_INET,socket.SOCK_DGRAM)
 sk.sendto(MESSAGE, (pkt[IP].dst, pkt[UDP].dport))
 packet.drop()

nfqueue = NetfilterQueue()
nfqueue.bind(2, encrypt)


def decrypt(packet):
 cipher_suite = Fernet(key)
 dec_vc=cipher_suite.decrypt(packet.get_payload())
 pkt = IP(packet.get_payload())
 MESSAGE = dec_vc
 sk=socket.socket(socket.AF_INET,socket.SOCK_DGRAM) 
 sk.sendto(MESSAGE,(pkt[IP].dst, pkt[UDP].dport))
 packet.drop()

nfqueue = NetfilterQueue()
nfqueue.bind(3, decrypt)
\end{lstlisting}
\end{table}

\begin{figure}[h]
	\begin{center}
		\includegraphics[scale=0.4]{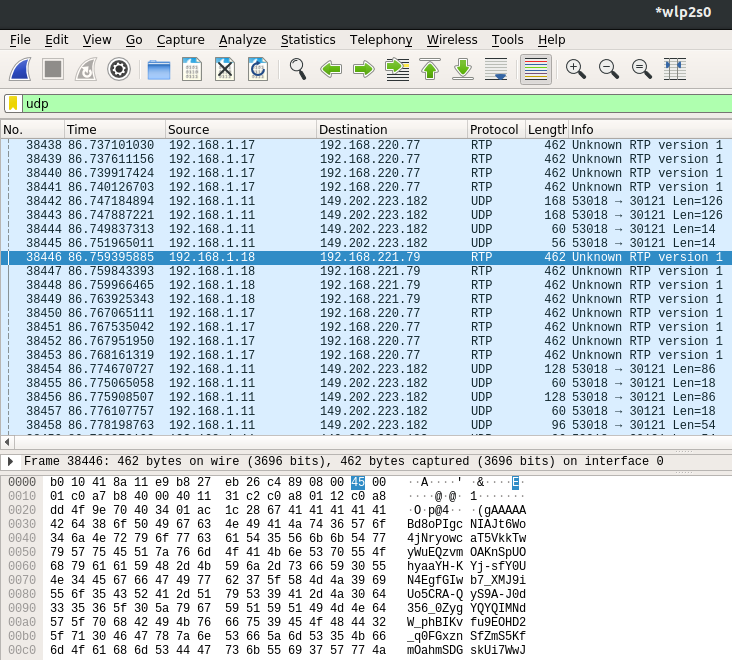}
	\end{center}
	\caption{A call sniffing attempt under our Phonejack 3 countermeasure}\label{fig:sniffenc}
\end{figure}

We can now launch a sniffing job through Wireshark. This simulates a MITM who actively attempts call sniffing between the two phones. What the attacker would intercept is nothing but encrypted RTP traffic, as shown in Figure \ref{fig:sniffenc}.

The decryption script is similar to the encryption script but with two differences. One is the method invoked on the payload, which, in this case is \textit{decrypt}. The other one is the Iptables queue that is accessed, in this case queue number 3.

\begin{figure}[h]
	\begin{center}
		\includegraphics[scale=0.25]{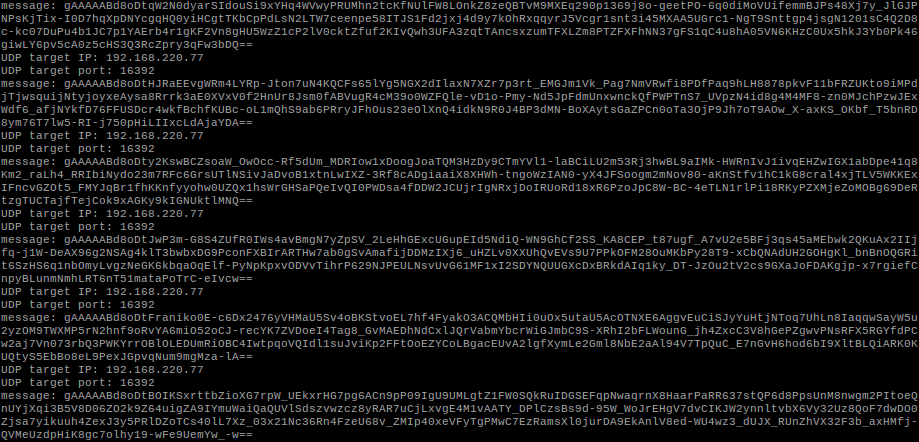}\\
		\vspace{0.25cm}
		\includegraphics[scale=0.25]{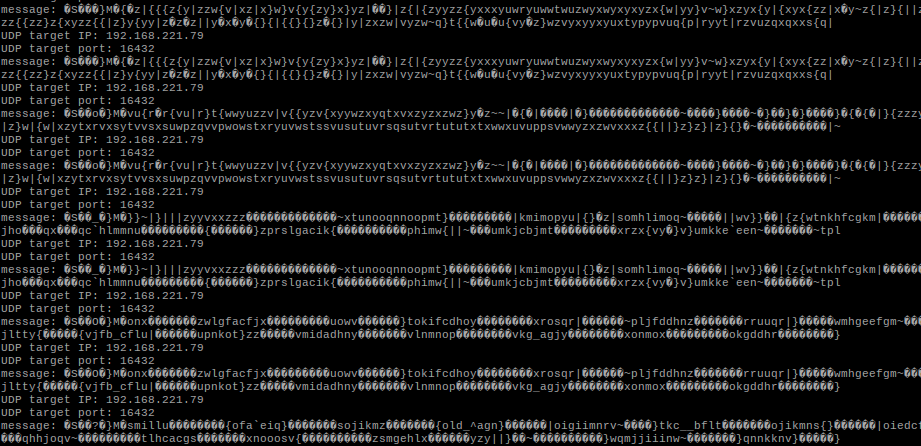}
	\end{center}
	\caption{Execution of our Phonejack 3 countermeasure on a Raspberry Pi}\label{fig:trafficoconPI}
\end{figure}

Figure \ref{fig:trafficoconPI} shows how a Raspberry Pi manages traffic under our Phonejack 3 countermeasure. The upper part of the Figure shows a terminal running the encryption routine, and hence displays an encrypted RTP stream; conversely, the lower part shows a terminal running the decryption routine, and hence displays unencrypted traffic.

\section{Related Work}\label{sec:related}
There is only room for a very brief treatment here. 

McGann and Sicker looked at security threats and tools for SIP-based VoIP technologies in 2005 \cite{Mcgann2005AnAO}. Their main conclusion was that testing tools do not always provide the coverage declared by the developers and may be difficult to install and configure properly. We find that work highly motivational but, regretfully, not followed up by much research. 

The most notable work is of 2010 by Keromytis \cite{angelostat}. He drew VoIP security statistics showing that 58\% of VoIP attacks are on Denial of Service, while 20\% are on eavesdropping, and hijacking. This reconfirms the importance of VoIP security measures and, in this vein, our work shows what can be done using modern freeware both for attack and defence purposes.

Next to the SIP-based VoIP technologies tackled in the present paper, also TLS-based solutions such as SIPS \cite{SIPS} and SRTP \cite{SRTP} must be mentioned, with the extra ``S'' in their acronyms arguably meaning some ``security''. In particular, SRTP uses asymmetric encryption to aim at authentication, integrity and confidentiality. It is clear that moving on to these technologies requires significant upgrades at both server and clients level. It is a matter of separate, future research to investigate how common the use of such technologies is today and whether it may suffer, in turn, exploitable weaknesses, perhaps under a socio-technical lens.

\section{Evaluations and conclusions}\label{sec:concl}
We have contended that VoIP and IoT are tightly intertwined. VoIP phones can be seen as early representatives as well as present enhancers of the IoT. Secure versions of VoIP protocols exist but are often neglected in favour of more traditional, unsecured technologies. This paper targets traditional VoIP, focusing on attacks and corresponding defence measures. The findings are demonstrated on common CISCO devices, for example costing in the region of \texteuro40 on Amazon Italy \cite{amazonit} and Amazon UK \cite{amazonuk} at the time of this writing.

Phonejack 1 conjectures that such devices may suffer vulnerabilities, documented or not, that could be exploited. Phonejack 2 demonstrates, also in a video clip \cite{videophonejack}, that an entire network of phones can be overwhelmed with ringing, cutting a vital institutional service and virtually making the target institution worth of evacuation due to noise levels. Phonejack 3 intercepts calls. It is reassuring that inexpensive Raspberry Pis can be programmed to counter the last two attacks, inviting a \textit{by design and by default} integration of such technology with the actual phones.

Our offensive and defensive experiments were conducted in an isolated environment but also inspired some information gathering at institutional level; the latter highlighted a weakness in the network separation in our Department. The finding was promptly reported to our IT team, ultimately resulting in the configuration of a stronger separation between the phone network and the computer network.

Our findings are significant. Although the worldwide diffusion of the devices in our testbed seems limited to a few hundreds from what can be discovered publicly, we argue that there are many more in use, correctly protected  by traditional network security measures such as VLANs and air gapping. Even so, from an outsider attacker standpoint, if a network attack point exists, then the outsider could leverage Phonejack attacks. From an insider attacker standpoint, network security measures could be as easy to bypass as to replace a phone with an attacking laptop.

A fundamental question we raised with printers \cite{overtrustprinter} firmly arises also in this case. Why are phones configured without any security measure at all when we are used to protecting our institutional laptops with a number of such measures, such as authentication, just to begin with? With the finally consolidated idea that our laptops host personal or sensitive data hence must be correspondingly protected \textit{even if} network security measures are in place, it is hard to justify why the same care is not devoted in practice to the verbal transmission of such data through voice calls. We cannot help but advocating that adequate security measures be wisely applied to \textit{every} computing node of the IoT.

\bibliographystyle{abbrv}
\bibliography{biblio}

\newpage
\appendix[A Primer on VoIP Protocols]\label{sec:introprotocols}
The essential protocols underlying VoIP are SIP and RTP.

\subsection{Session Initiation Protocol (SIP)}\label{subsec:sip}
Developed by the Internet Engineering Task Force (IETF), SIP consists in a telephone signaling protocol used to establish, modify and conclude VoIP phone calls \cite{SIP}. More precisely, it uses the UDP transport protocol with default port 5060 and has the following functions: i) authenticate, locate and acquire the audio coding preferences of the clients; ii) invite clients to participate in a session; iii) establish session connections; iv) carry a description of the session; v) manage any changes to the session parameters; vi) conclude telephone sessions.

The SIP protocol is based on a Client-Server system, in fact, generally there is a dedicated machine that plays this role, called SIP server. A client after being authenticated by the SIP server, can establish a connection with another client by the following steps:
\begin{enumerate}
	\item  The client sends this invitation message to the server. With this message, the client asks the server to establish a connection with the client indicated in the \texttt{user} parameter.\\
	\texttt{SIP INVITE:\textless sip : user@serverhost\textgreater}
	
	\item The server responds with a ``Trying" message to the client.
	
	\item The server forwards the invitation of the first phase to the recipient.
	
	\item The receiver sends a ``Trying" message to the server.
	
	\item The recipient sends a ``Ringing" message to the server, then, the server sends the same message to the sender. This message generates the typical call tone.
	
	\item When the recipient accepts the call, it sends an ``OK" message to the server. Subsequently, the server will notify the sender of the acceptance of the call.
	
	\item After the reception of the ACK message relating to the sixth phase, communication and transmission of the audio signal through the RTP protocol is established.
	
	\item To conclude the call, each party sends a ``BYE" message to the server, which will inform the other party.
\end{enumerate}
\subsection{Real-time Transport Protocol (RTP)}\label{subsec:rtp}
Also developed by the IETF, the Real-time Transport Protocol (RTP) complements SIP by providing end-to-end network transport functions suitable for real-time applications such as VoIP \cite{RTP}. RTP also uses UDP at the transport level. The RTP protocol does not have a default port, so different VoIP applications may choose a different port number. For example, Cisco SPA devices select the port number randomly from an established range of ports at each call. Once the RTP session is established, the clients use the audio coding specifications previously established thanks to the SIP protocol.

The information provided by the RTP protocol includes timestamps for synchronization, sequence numbers useful in case of packet loss and the payload format that indicates the coded format of the data. Thanks to these fields in the RTP packets, the call can be reconstructed via Wireshark.
Unfortunately, the RTP protocol has some limitations, RTP does not control the quality of service (QoS), does not guarantee the delivery of packets and does not provide automatic retransmission of packets in case of loss.

\end{document}